\documentclass[aps, prd, twocolumn, showpacs, superscriptaddress, groupedaddress]{revtex4-1} 

\usepackage{graphicx}	
\usepackage{amssymb}
\usepackage{dcolumn}
\usepackage{color}
\usepackage{subfigure, rotating, bm, array}
\usepackage{mathtools}
\usepackage{amsmath}
\usepackage{xcolor}
\usepackage[pagebackref=false, colorlinks=true]{hyperref}
\hypersetup{
linkcolor=blue,     
citecolor=blue,     
urlcolor=blue} 

\begin{document}

\title{Light trajectory and shadow shape in the rotating naked singularity}
\author{Vishva Patel}   
\affiliation{International Center for Cosmology, Charusat University, Anand, GUJ 388421, India}
\author{Divya Tahelyani}
\email{tahelyanidivya118@gmail.com}
\affiliation{International Center for Cosmology, Charusat University, Anand, GUJ 388421, India}
\author{Ashok B. Joshi}
\email{gen.rel.joshi@gmail.com}
\affiliation{International Center for Cosmology, Charusat University, Anand, GUJ 388421,  India}
\author{Dipanjan Dey}
\email{deydipanjan7@gmail.com}
\affiliation{International Center for Cosmology, Charusat University, Anand, GUJ 388421, India}
\author{Pankaj S. Joshi}
\email{psjcosmos@gmail.com}
\affiliation{International Centre for Space and Cosmology, Ahmedabad University, Ahmedabad, GUJ 380009, India\\
International Center for Cosmology, Charusat University, Anand, GUJ 388421,  India}
\date{\today}
\begin{abstract}
In this paper, we investigate the light trajectories and shadow properties in the rotating version of null naked singularity (NNS) spacetime which is derived using the Newman- Janis algorithm without complexification method. We discuss some of the geometrical properties and causal structure of Rotating Naked Singularity (RNS) spacetime. The gravitational lensing in a rotating naked singularity is analyzed, and the results are compared to those of a Kerr black hole. In the case of a Kerr black hole, the photon sphere exists for both prograde and retrograde photon orbits, whereas for RNS, the photon sphere exists only for retrograde photon orbits. As a result, the naked singularity projects an arc-shaped shadow that differs from the contour-shaped shadow cast by a Kerr black hole.

\end{abstract} 
\maketitle
\section{Introduction}
So far, Einstein's theory of general relativity is a remarkably successful theory of gravity.\,\,When large enough masses collapse under the pull of their own gravity, the general theory of relativity predicts that a spacetime singularity would develop.\,\,Under certain conditions, a singularity theorem demonstrates the inevitability of the emergence of a spacetime singularity.\,\,However, there was no such theorem about the existence of an event horizon enclosing the singularity.\,\,In 1969, Roger Penrose introduced the cosmic censorship conjecture (CCC), which asserts that horizon-less strong spacetime singularities are not feasible \cite{Penrose:1964wq}. Nonetheless, there are several studies on the continuous gravitational collapse of the inhomogeneous matter cloud that illustrates that spacetime singularities developed during gravitational collapse can be seen by an outside observer \cite{joshi,goswami,mosani1,mosani2,mosani3,mosani4,Deshingkar:1998ge,Jhingan:2014gpa}. The visibility of spacetime singularity depends upon the initial conditions of the collapsing matter \cite{Joshi:2011zm}. In the regime of general relativity, such a scenario can also arise where the continual gravitational collapse may lead to an equilibrium static spacetime with a central visible or naked singularity, e.g., Joshi-Malafarina-Narayan (JMN) spacetime~\cite{Joshi:2011zm, JNW,Joshi:2020tlq}.

It is evident from the above diacussion that the naked singularity can arise as the final state of continuous gravitational collapse of an inhomogeneous matter cloud. Therefore, if naked singularity exists in nature, it would have distinct physical traces. Photon trajectories may extract information about the distinguish spacetime structure near a naked singularity. Therefore, detailed study of null geodesics near naked singularity is very much important to predict the physical signature of naked singularity. 
 The route is offered by gravitational lensing by naked singularities, as described by several authors in recent literature \cite{Virbhadra:2002ju,Sahu:2012,Paul:2020fc,Dey:2013ya,Dey:2020hf,Shaikh:2019lcc}.
Gravitational lensing is defined as both the deflection of light and the variation in visual brightness of the radiating source induced by a gravitational field. The study of analyzing the shadow properties of a naked singularity is a curious and useful technique for identifying the spacetime structure of naked singularity. The observation of the shadow also gives an initial way of determining the properties of the naked singularity. In 2019, the Event Horizon Telescope (EHT) group released the shadow image of a compact object located at the center of Messier 87 (M87) galaxy \cite{EventHorizonTelescope:2019dse}. Besides that, recently the EHT group released the shadow image of a compact object located at the center of Milky way galaxy Sgr $A^{*}$  \cite{EventHorizonTelescope:2022xnr,EventHorizonTelescope:2022tzy,EventHorizonTelescope:2022ago,EventHorizonTelescope:2022vjs,EventHorizonTelescope:2022wok,EventHorizonTelescope:2022urf,EventHorizonTelescope:2022xqj,EventHorizonTelescope:2022xqj,EventHorizonTelescope:2022exc,EventHorizonTelescope:2022gsd}. They explore the shadow image cast by various models and compared it to the shadow image of Sgr $A^{*}$, concluding that the JMN-1 naked singularity is one of the best possible non-spinning black hole mimickers~\cite{EventHorizonTelescope:2022xqj}. Furthermore, they suggest that the metric tests cannot rule out the possibility of the naked singularities as a Sgr $A^{*}$ compact object~\cite{EventHorizonTelescope:2022xqj}. These observations have opened up new possibilities for analysing various naked singularity spacetimes as black hole mimickers. In literature, many researchers probe into the shadow cast by other types of compact objects such as gravastar, regular black holes, and wormholes \cite{Gralla:2019xty,Abdikamalov:2019ztb,Vagnozzi:2019apd,Gyulchev:2019tvk,Dey:2013yga,Dey+15,Dey:2020haf,atamurotov_2015, abdujabbarov_2015b, ohgami_2015, stuchlik_2019, Kaur, Sakai, Shaikh:2021,Li:2021,Hu:2020usx}. It is now known that event horizon and photon sphere are not necessary for a shadow formation. In papers~\cite{Joshi:2020tlq}, authors show that existence of upper bound of the effective potential of null geodesic is the necessary and sufficient condition for shadow formation.

Since every compact object in our universe possesses inherent angular momentum, a rotating counterpart of the singular static spacetime is essential for a more realistic scenario. In this work, we first construct a rotating version of null naked singularity (NNS) spacetime to explore its physical features. The rotating naked singularity spacetime could be obtained by applying the Newmann-Janis Algorithm (NJA) to the static NNS. Furthermore, while applying the NJA to other spacetime metrics, such as naked singularities and other black hole solutions, the final result in Eddington-Finkelstein coordinates (EFC) may not be transformed into the Boyer-Lindquist coordinates (BLC) due to the complexification process mentioned in \cite{Mustapha2011}. In \cite{Mustapha2011}, Mustapha Azreg-Anou show how the rotating spacetime obtained by the NJA and written in the EFC could be transformed into the BLC by skipping the complexification process. Recently, the rotating form of the polytropic black hole and Janis-Newmann-Winicour naked singularity spacetimes have been generated using the NJA without complexification \cite{Solanki:2021mkt}.
In this paper, we apply the NJA without complexification to set the rotating counterpart of static NNS spacetime discussed in~\cite{Joshi:2020tlq,Paul:2020fc,Dey:2020bgo}. We analyse the properties of light geodesics and the shadow around RNS and compare the results with those of the Kerr black hole. The results reveal that the geometry of the rotating version of NNS spacetime has significant different features in terms of the light trajectories and shadow shape than the Kerr black hole, which provides an effective tool for distinguishing the RNS from the Kerr black hole.

This paper is organised as follows. In Section~(\ref{sec2}), we obtain the rotating naked singularity (RNS) solution by applying the NJA to static NNS spacetime. In Section~(\ref{sec3}), we investigate whether the energy conditions are satisfied in RNS spacetime. In Section~(\ref{sec4}), we compare the light trajectory in the Kerr black hole and the RNS spacetime. We then compare the shadow shape for the Kerr black hole and the RNS spacetime. Ultimately in Section~(\ref{sec_discussion}), we conclude and discuss the outcome of the work. Throughout the paper, we consider geometrized units, where the gravitational constant(G) and the speed of light(c) are set equal to one. The signature of the metric is (-,+,+,+).

\section{Newman-Janis Algorithm without Complexification}\label{sec2}

To apply the Newman- Janis algorithm without complexification, let us consider a general form of spherically symmetric and static spacetime, in which, the $g_{tt}(r)$ and $g_{rr}(r)$ are inversely proportional to each other.\,The spacetime metric is,
\begin{equation}
ds^2=-A(r)dt^2+ \frac{dr^2}{B(r)} + C(r)d\Omega^2.
\label{general_metric}
\end{equation}
where, $d\Omega^2 = d\theta^2 + \sin^2\theta d\phi^2$. The $g_{tt}(r)$ term is $A(r)$, $g_{rr}(r)$ term is $B(r)$ and $C(r) = r^2$. 
One can use the null coordinates $\{u,r,\theta,\phi\}$ to describe the above spacetime. The coordinate transformation can be written as,
\begin{equation}
du=dt-\frac{dr}{\sqrt{A(r) B(r)}},
\end{equation}
and in that null coordinates, the spacetime metric (\ref{general_metric}) takes the following form,
\begin{equation}
ds^2=-A(r)du^2-2\sqrt{\frac{A(r)}{B(r)}}dudr + C(r)d\Omega^2.
\label{general_metric_null}
\end{equation}
One can write the null tetrads of the above spacetime metric as,
\begin{eqnarray}
&& l^\mu=\delta^\mu_1\,, \label{l_mu} \\
&& n^\mu=\sqrt{\frac{B(r)}{A(r)}}\delta^\mu_0-\frac{1}{2}B(r)\delta^\mu_1\,, \label{n_mu} \\
&& m^\mu=\frac{1}{\sqrt{2C(r)}} \left(\delta^\mu_2+\frac{i}{\sin\theta}\delta^\mu_3 \right) \label{m_mu}
\end{eqnarray}
The complexification procedure of the NJA extrapolates the function (i.e., $A(r)$, $B(r)$, $C(r))$ as the real function of the radial coordinate r and its complex conjugate $ \bar{r}$. The solution is to skip this step and move on to the next, which requires complex coordinate transformations. In the complex coordinate transformation the spacetime metric is derived in the null coordinates $\{u,r,\theta,\Phi\}$. One can transform this into the Boyer-Lindquist coordinates~(BLC) $\{t,r,\theta,\phi\}$ using the transformation,
\begin{eqnarray}
&& du=dt-\lambda(r)dr
\label{du new}\\
&& d\Phi=d\phi-\chi(r)dr
\label{dphi new}
\end{eqnarray}
 where we get the BLC transformation function $\chi$ and $\lambda$ as,   
\begin{eqnarray}
&& \chi(r)=\frac{a}{ CB +a^2},
\label{chi}\\
&& \lambda(r)=\frac{K+a^2}{CB+a^2},
\label{lambda}
\end{eqnarray}
where, $K=C\sqrt{\frac{B}{A}}$. In the Boyer- Lindquist coordinates ($t,r,\theta,\phi$), the metric can be represented as, 

\begin{eqnarray}
  ds^2 = -\left(1-\frac{2f}{\rho^2} \right)dt^2 + \frac{\rho^2}{\Delta} dr^2 + \rho^2 d\theta^2 + \frac{\Sigma\sin^2\theta}{\rho^2} d\phi^2 \nonumber \\ -\,  \frac{4af\sin^2\theta}{\rho^2} dt d\phi. \,\,\,\,\,\,\,\,\,\,
  \label{general_rotating_metric}
\end{eqnarray}
Where,
\begin{eqnarray}
   && \rho^2 = K(r) + a^2\cos^2\theta\,, \\
   && f = \frac{K(r)-C(r)B(r)}{2}\,, \\
  && \Delta = C(r) B(r) + a^2\,, \\
   && \Sigma = \big(K(r)+a^2 \big)^2 - a^2\Delta\sin^2\theta.
\end{eqnarray}
The rotating solution (\ref{general_rotating_metric}) reduces to the static when $a \to0$, and spherically symmetric spacetime metric if $A(r)$ and $B(r)$ are inversely proportional to each other.

\section{Rotating compact object}\label{sec3}
In this section, we present a new axially symmetric, asymptotically flat RNS spacetime. We first discuss the basic spacetime properties of this RNS spacetime, and then we briefly discuss the Kerr black hole spacetime.
\subsection{Rotating Naked Singularity (RNS) Spacetime}
In the section (\ref{sec2}), we develop the rotating form of static and spherically symmetric metric using Janis- Newman algorithm without complexification. In \cite{Joshi2020}, authors describe a null-naked singularity spacetime which resembles with Schwarzschild spacetime far away from the center. The metric of the null-naked singularity spacetime is,
\begin{equation}
    ds^2 = -\frac{1}{\left(1 + \frac{M}{r}\right)^2} dt^2 + \left(1 + \frac{M}{r}\right)^2 dr^2 + r^2 d\Omega^2\,\, ,
    \label{null}
\end{equation}
where $d\Omega^2 = d\theta^2 + \sin^2\theta d\phi^2$ and $M$ is the Arnowitt- Deser-Misner (ADM) mass of the above spacetime metric. One can write down the following rotating counterpart of the above mentioned static NNS spacetime using the NJA without complexification technique described in the previous section,
\begin{widetext}
\begin{equation}
    ds^2 = -\left(1-\frac{2f}{\rho^2} \right)dt^2 - \frac{4af\sin^2\theta}{\rho^2} dt d\phi + \frac{\Sigma\sin^2\theta}{\rho^2} d\phi^2 + \frac{\rho^2}{\Delta} dr^2 + \rho^2 d\theta^2, \\  
    \label{null rotating}
    \end{equation}
\end{widetext}
where, 
\begin{eqnarray}
    && f = \frac{r^2M^2+2Mr^3}{2(M+r)^2}\,, \\
    && \rho^2 = r^2 + a^2\cos^2\theta\,, \\
    && \Delta =  \frac{r^2}{\left(1+\frac{M}{r}\right)^2} + a^2\,, \\
    && \Sigma = (r^2 + a^2)^2 - a^2\Delta\sin^2\theta.
\end{eqnarray}

Here, we note that the above rotating metric will reduce to the Eq.\,(\ref{null}) in the limit of $a\to0$. Now,  we must proceed to verify the validity of the energy conditions in this rotating spacetime. As we know, the energy-momentum tensor can be written as,
\begin{equation}
     T^{\mu\nu} = \rho_e u^\mu u^\nu + P_r e_r^\mu e_r^\nu + P_{\theta} e_\theta^\mu e_\theta^\nu +P_{\phi}  e_\phi^\mu e_\phi^\nu\,\, ,
\end{equation}
where $\rho_e$ is the energy density, and $P_i$ $(i\to r,\theta,\phi)$ are the main pressure components. Using the Einstein field equation $G_{\mu\nu}=T_{\mu\nu}$, we can write down the $\rho_e$ and $P_i$ as,
\begin{eqnarray}
    && \rho_e = G_{\mu\nu}\, u^\mu u^\nu  \label{rho general}\,, \\
    && P_r = G_{\mu\nu}\, e_r^\mu e_r^\nu  = g^{rr} G_{rr} \label{Pr general}\,, \\
    && P_\theta = G_{\mu\nu}\, e_\theta^\mu e_\theta^\nu = g^{\theta\theta}\, G_{\theta\theta} \label{Ptheta general}\,, \\
    && P_\phi = G_{\mu\nu}\, e_\phi^\mu e_\phi^\nu  \label{Pphi general}.
\end{eqnarray}
\begin{figure}[ht!]
\centering
{\includegraphics[width=85mm]{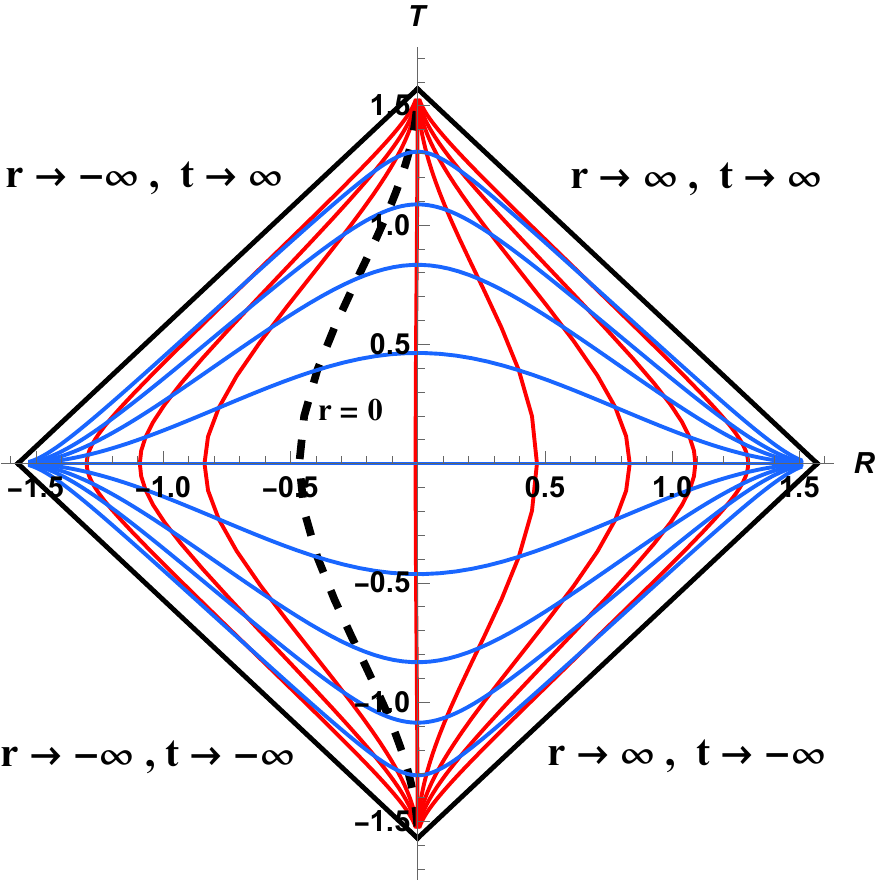}
 \caption{Penrose diagram of rotating naked singularity }
 \label{null_penrose}}
 \end{figure}
From (\ref{rho general}), (\ref{Pr general}), (\ref{Ptheta general}) and (\ref{Pphi general}), the components of the energy-momentum tensor of the
RNS spacetime are as follows (considering $\theta = \pi/2$),
\begin{eqnarray}
    && \rho_e = \frac{M^2 (M+3 r)}{r^2 \left(M+r\right)^3}\,, \\
    && P_r = -\frac{M^2 (M+3 r)}{r^2 \left(M+r\right)^3} \,, \\
    && P_\theta = P_\phi = \frac{3 M^2}{(M+r)^4}.
\end{eqnarray}
From the above expressions, it can be verified that the RNS spacetime~(\ref{null}) satisfies the weak energy condition ($\rho_e \geq 0$, $\rho_e+P_i \geq 0$), strong energy condition ($\rho_e+\sum P_i \geq 0$, $\rho_e+P_i \geq 0$), the dominant energy condition ($\rho_e \geq |P_i|$), and the null energy condition ($\rho_e+P_i \geq 0$) considering $\theta = \frac{\pi}{2}$. The presence of horizons in a stationary spacetime geometry are generally given by the condition $g_{rr}\to\infty$. Equivalently, the null values of $\Delta(r)$ represent the locations of horizons. From $\Delta(r)=0$, we find that, the radius of horizons becomes imaginary for RNS spacetime, which indicates that it is a horizon-less compact object.
The Kretschmann scalar which is defined as, $k = R_{\alpha \beta \gamma \delta}\, R^{\alpha \beta \gamma \delta}$ becomes infinite in the RNS spacetime
for $ r \to 0 $ and $\theta \to \frac{\pi}{2}$. This behaviour of the Kretschmann scalar indicates the presence of a ring singularity in the equatorial plane with a coordinate radius $a$. The causal structure of the RNS spacetime can be examined using the Penrose diagram. To generate the Penrose diagram, we transform the $t$ and $r$ coordinates into $T$ and $R$ coordinates, respectively, by using the following transformations,
\begin{eqnarray}
    T = \frac{1}{2}(\tan^{-1} (t + r^{*}) + \tan^{-1} (t - r^{*}))\,,\\
    R = \frac{1}{2}(\tan^{-1} (t + r^{*}) - \tan^{-1} (t - r^{*}) )\,.
\end{eqnarray}
where $r^{*} = \int \left(\frac{g_{rr}}{g_{tt}}\right)^{\frac{1}{2}} dr$. 
The new coordinates $T$ and $R$ are used to describe the entire spacetime manifold in a specific compactified diagram. We need to check the values of $T$ and $R$ at $r \to 0$ to see if the spacetime singularity is time like or null like. For the finite value of $t$ and $r \to 0$, if $r^* \to -\infty$ then one can conclude that the singularity at $r=0$ is nulllike, since $\lim_{r^* \to -\infty}R=-\pi\,\, ,\lim_{r^* \to -\infty}T=0$. On the other hand, for the finite value of $t$ and $r \to 0$, if the singularity exists at a constant $r$, then the singularity is timelike. From the Penrose diagram in Fig.~(\ref{null_penrose}), one can see that the singularity in the RNS spacetime is a timelike singularity.

\subsection{Kerr Black Hole Spacetime}
In the Boyer- Lindquist coordinates $(t, r, \theta, \phi)$, the  line element of a stationary, axisymmeric, and rotating Kerr spacetime is given by Eq.\,(\ref{null rotating}) with, 
\begin{eqnarray}
&&a=\frac{J}{M}\,,\\
&&f= M r\,,\\
&&\rho^2 = r^2+ a^2 \cos^2\theta\,,\\
&&\Delta = r^2-2 M r +a^2\,,\\
&&\Sigma = (r^2+a^2)^2-a^2\Delta \sin^2\theta\,.
\label{parameters}
\end{eqnarray}
The spin parameter $a$ has the dimensions of length in geometrized units. As discussed above, the positions of horizons in Kerr geometry can be determined using the condition $\Delta(r)=0$, which gives
\begin{equation}
r_\pm= M\pm\sqrt{M^2-a^2}\,.
\end{equation}
The surface at $r=r_-$ represents a Cauchy horizon, while the surface at $r=r_+$ represents the event horizon. In the extreme case $a=M$, both the horizons coincide at $r=M$. For $a>M$, the horizons become imaginary, and the singularity in the Kerr black hole becomes naked.

\section{LIGHT TRAJECTORIES IN ROTATING NAKED SINGULARITY AND KERR BLACK HOLE}\label{sec4}

A general rotating spacetime in the Boyer-Lindquist co-ordinates can be given by,
\begin{equation}
    ds^2=-g_{tt}dt^2+g_{rr}dr^2+g_{\theta\theta}d\theta^2+g_{\phi\phi}d\phi^2 -2g_{t\phi}dt d\phi\,.\label{line_element}
\end{equation}
In order to obtain the shadow of these compact objects, it is required to study the null geodesics in the spacetimes. The null geodesics can be derived from the Lagrangian given by
\begin{equation}
    2\mathcal{L}=g_{\mu\nu} \dot{x}^{\mu} \dot{x}^{\nu}\,,\label{L}
\end{equation}
where the overdot represents the differentiation with respect to the affine parameter ($\lambda$) along the light geodesic. For the line element given in Eq.\,(\ref{line_element}),  Eq.\,(\ref{L}) becomes
\begin{equation}
2\mathcal{L}=- g_{tt} \dot{t}^2 + g_{rr} \dot{r}^2 + g_{\theta \theta} \dot{\theta}^2 +g_{\phi \phi}\dot{\phi}^2- 2g_{t\phi}\dot{t}\dot{\phi}\,.\label{2L}
\end{equation} Setting $\theta=\pi/2$ for equatorial plane in  RNS~(Eq.~(\ref{null rotating})) and Kerr black hole~(Eq.~(\ref{parameters})) metric, the above equation yields,
\begin{widetext}
	\begin{eqnarray}
2\mathcal{L}_{RNS}& =& - \frac{1}{(1+\frac{M}{r})^2} \dot{t}^2 + \frac{r^2}{\Delta } \dot{r}^2 + \frac{\left(a^2+r^2\right)^2-a^2 \Delta }{r^2}\dot{\phi}^2-\frac{2 a M (M+2 r)}{(M+r)^2}\dot{t}\dot{\phi}\,, \label{L_null}
\\
2\mathcal{L}_{Kerr}& =& -\left(1-\frac{2M}{r}\right) \dot{t}^2 + \frac{r^2}{\Delta} \dot{r}^2 + \left(r^2 +a^2 +\frac{2M a^2}{r}\right) \dot{\phi}^2 - \frac{4a M }{r}\dot{t}\dot{\phi}\,.\label{L_Kerr}
	\end{eqnarray}
\end{widetext}
Since the Lagrangian is independent of $t$ and $\phi$ co-ordiantes, $p_t$ and $p_{\phi}$ are conserved along the light geodesic.  From the above Lagrangian, the $t$- and $\phi$- component of momenta can be deduced:
\begin{eqnarray}
&&p_{t} = \frac{\partial\mathcal{L}}{\partial \dot{t}}= - g_{tt}\dot{t}-g_{t\phi}\dot{\phi}= -\epsilon\,,\label{pt}
\\
&&p_{\phi} = \frac{\partial\mathcal{L}}{\partial \dot{\phi}} =  - g_{t\phi}\dot{t} + g_{\phi\phi}\dot{\phi} = L\,,\label{pphi}
\end{eqnarray}
where $\epsilon$ and $L$ are, respectively, the conserved energy and angular momentum (about the axis of symmetry) of the photon evaluated at spatial infinity. The $r$- component of the momentum is given by
\begin{equation}
p_{r} = \frac{\partial\mathcal{L}}{\partial \dot{r}} = g_{rr} \dot{r}\,.\label{pr}
\end{equation}
Solving Eq.\,(\ref{pt}) and Eq.\,(\ref{pphi}) for $\dot{t}$ and $\dot{\phi}$, we arrive at
\begin{eqnarray}
&&\dot{t}=\frac{-L g_{t\phi} +\epsilon g_{\phi\phi}}{g_{tt}g_{\phi\phi}+g^2_{t\phi}}\,,\label{tdot}\\
&&\dot{\phi}=\frac{L g_{tt} + \epsilon g_{t\phi} }{g_{tt}g_{\phi\phi}+g^2_{t\phi}}\,.\label{phidot}
\end{eqnarray}
Four velocity $\dot{x}^{\mu}$ in Eq.\,(\ref{L}) is normalized to zero  for null geodesics. Hence, for photons we have $\mathcal{L}=0$. Inserting the solutions Eq.\,(\ref{tdot}) and Eq.\,(\ref{phidot}) in  Eq.\,(\ref{2L}), we obtain
\begin{equation}
\dot{r}^2 = \frac{-g_{tt}L^2- 2 \epsilon L g_{t\phi}+\epsilon^2 g_{\phi\phi}}{g_{rr}(g_{tt} g_{\phi\phi} + g^2_{\phi\phi})}\,. \label{rdotsq}
\end{equation}
The shape of the photon orbit can be described by how $\phi$ changes with $r$. Using Eq.\,(\ref{phidot}) and Eq.\,(\ref{rdotsq}), we find
\begin{equation}
\frac{d\phi}{dr}=\frac{(L g_{tt}+ \epsilon g_{t\phi})\sqrt{g_{rr}}}{\sqrt{(g_{tt} g_{\phi \phi}+g^2_{t\phi}) (-g_{tt} L^2 - 2 \epsilon L g_{t \phi}+\epsilon^2 g_{\phi\phi})}}
\end{equation}For the simplicity, we define a new variable in above equation such that $u = 1/r$. Now, we can write down the second order differentiation of $u$ with respect to  $\phi$ in the following way,
\begin{equation}
	\frac{d^2 u}{d{\phi}^2}= \frac{2}{r^3}\left(\frac{dr}{d\phi}\right)^2-\frac{1}{2 r^2}\frac{d}{dr}\left(\frac{dr}{d\phi}\right)^2\label{orbit_eq}\,.
\end{equation}
We can construct the orbit equation in the RNS and Kerr spacetimes by employing the corresponding expression of $\frac{dr}{d\phi}$ and the components of the metric tensor of the spacetimes in the Eq.\,(\ref{orbit_eq}). The trajectories in the rotating equatorial geometry will differ depending on whether the particle or photon rotates about the axis of symmetry in the same direction (co-rotating or prograde) as the rotating compact object or in the opposite direction (counter-rotating or retrograde). By numerically solving the orbit equations and plotting the resultant light trajectories, we can easily visualise how light is lensed in the RNS and Kerr spacetime. The behaviour of light trajectories and the appearance of shadow shapes in RNS and Kerr black hole are discussed in the next section.\\
\section{Equatorial Photon Orbits and Shadow Outline}
 In the general theory of relativity, the Hamilton-Jacobi formalism is favourable for uncovering the nature of the shadow. The action in the Hamilton-Jacobi equation is given by,
\begin{equation}
    2 \frac{\partial S}{\partial \lambda} = - g^{\mu \nu} \frac{\partial S}{\partial x^{\mu}} \frac{\partial S}{\partial x^{\nu}},
    \label{Hamil Jac}
\end{equation}
where $S$ is the Jacobi action, also known as Hamilton's principal function, $\lambda$ is an affine parameter, and $g^{\mu \nu}$ is the metric tensor of the geometry. If the above differential Eq.\,(\ref{Hamil Jac}) has a separable solution, then in terms of the known constants of the motion, it should take the form,
\begin{equation}
    S = \frac{1}{2} \mu^{2} \lambda - \epsilon t + L \phi + S_r(r) + S
    _\theta (\theta)\,.
    \label{action}
\end{equation}
\begin{figure*}[ht!]
\centering
\subfigure[\, For an rotating naked singularity ]
{\includegraphics[width=85mm]{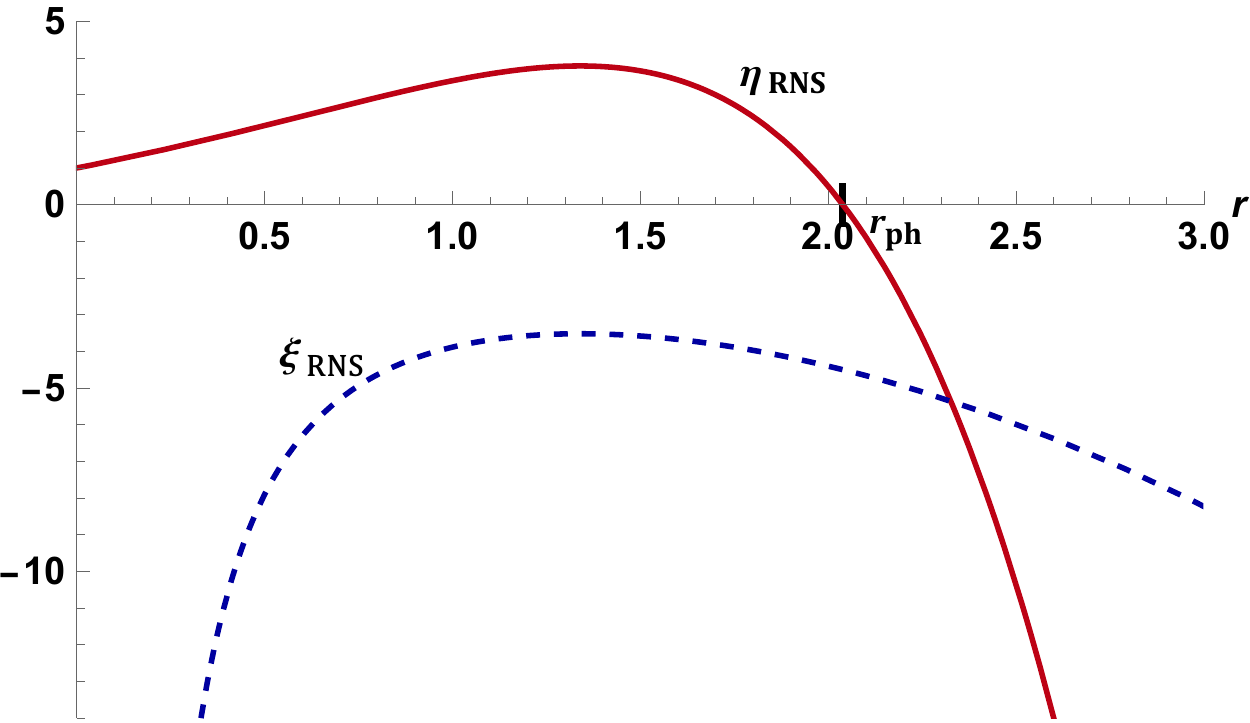}\label{fig:eta_xi_null}}
\hspace{0.5cm}
\subfigure[\, For a Kerr black hole]
{\includegraphics[width=85mm]{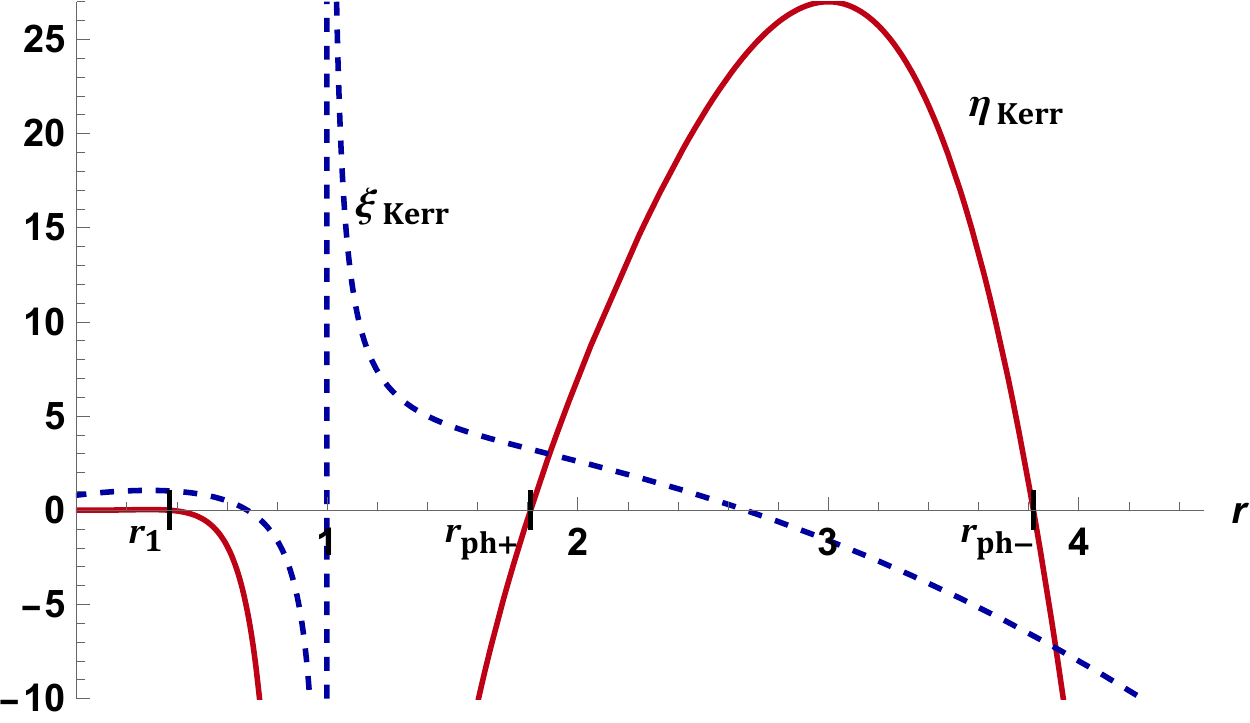}\label{fig:eta_xi_kerr}}
 \caption{Fig. (\ref{fig:eta_xi_null}) and (\ref{fig:eta_xi_kerr}) demonstrate the behaviour of $\eta$ and $\xi$ for the RNS and Kerr black hole as a function of radial coordinate $r$. Here, we consider $a=0.8$ and $M=1$.}\label{fig:V}
\end{figure*}
Here, $\mu$ is the rest mass of the test particle, and $\mu = 0$ for null geodesic. The functions $S_r(r)$ and $S_\theta(\theta)$ are variable ($r$ and $\theta$) specified functions.

 Substituting Eq.\,(\ref{action}) into Eq.\,(\ref{Hamil Jac}) and using the contravariant metric tensor components, we obtain the following equation,
\begin{widetext}
    \begin{equation}
    (L^2 cosec^2\theta - a^2 \epsilon^2) cos^2\theta - \left(\frac{d S_\theta(\theta)}{d \theta} \right)^2 = \frac{1}{\Delta} ((r^2 + a^2)\epsilon - a L)^2 - (L - a \epsilon)^2 + \Delta \left(\frac{d S_r(r)}{d r}\right)^2\,.
    \label{r_theta_combined}
    \end{equation}
\end{widetext}
Here, the equation above is the only function of $\theta$ and $r$. Therefore, one can separate both sides of Eq.~(\ref{r_theta_combined}) using the separation constant as,
\begin{equation}
        (L^2 cosec^2\theta - a^2 \epsilon^2) cos^2\theta - \left(\frac{d S_\theta(\theta)}{d \theta} \right)^2 = K,  
        \label{theta part}
\end{equation}

\begin{figure*}[ht!]
\centering
\subfigure[Prograde Light trajectories in the Kerr Black hole for $M=1, a=0.8$]
{\includegraphics[width=80mm]{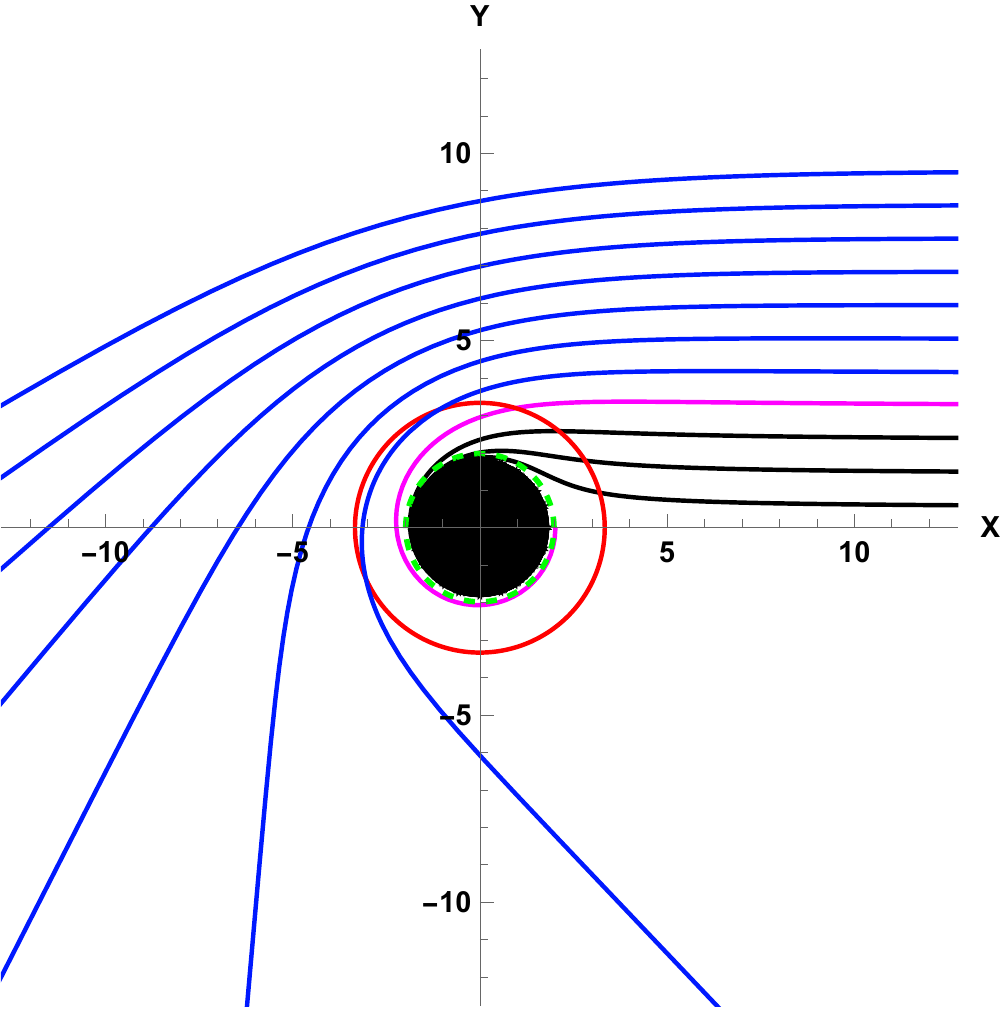}\label{fig:lensing1}}
\hspace{1cm}
\subfigure[Retrograde light trajectories in the Kerr black hole for $M=1, a=0.8$]
{\includegraphics[width=80mm]{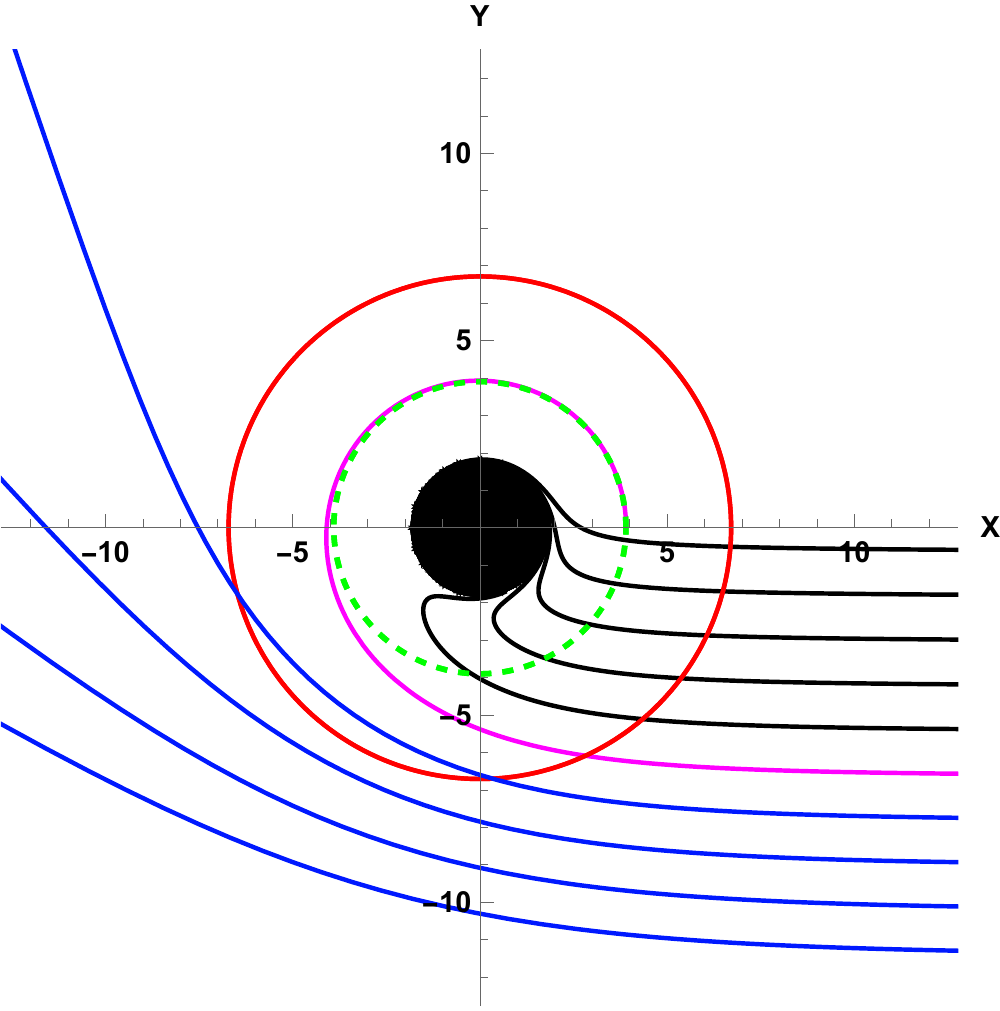}\label{fig:lensing2}}
\subfigure[Prograde Light trajectories in the rotating naked singularity for $M=1, a=0.8$]
{\includegraphics[width=80mm]{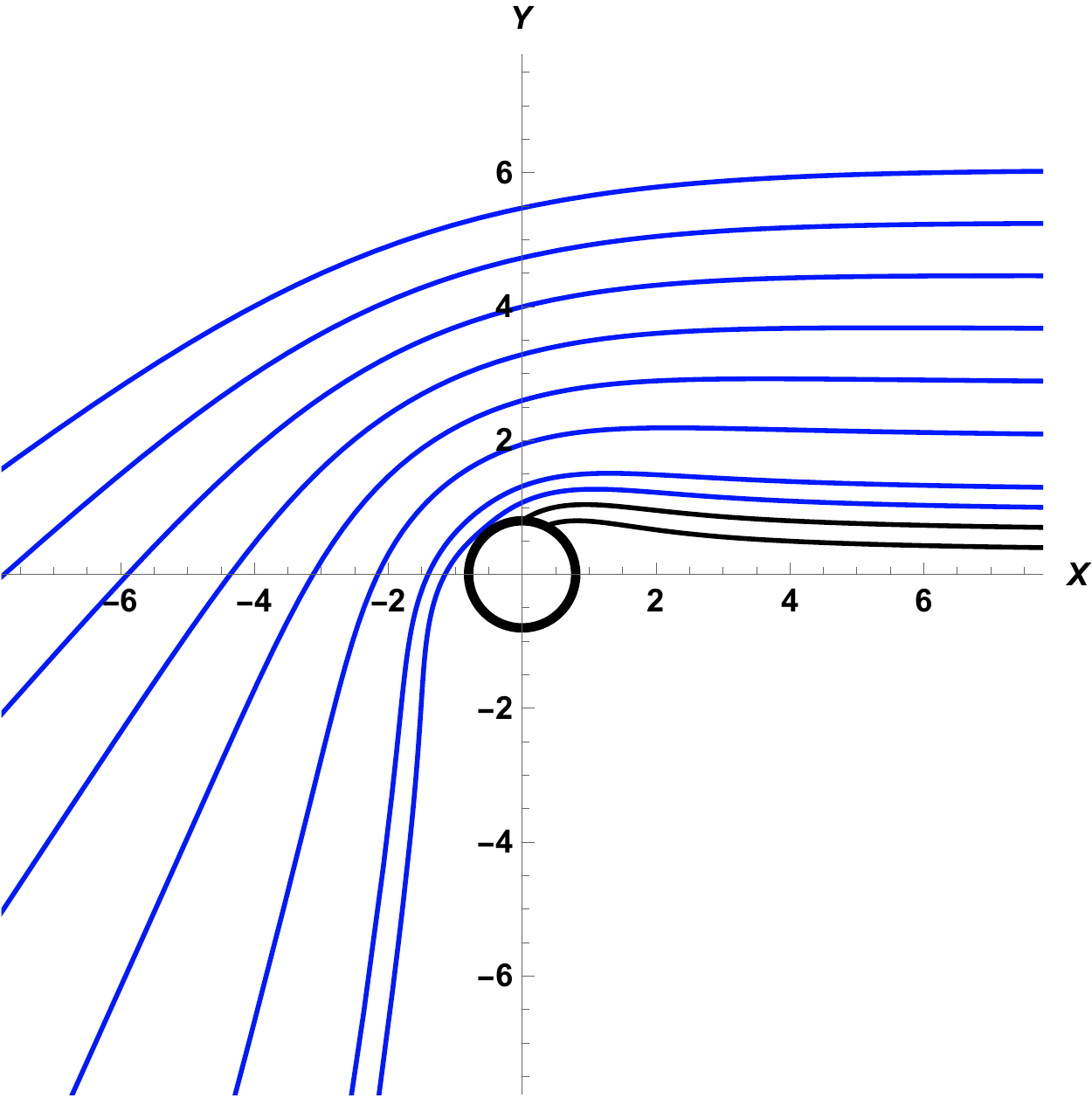}\label{fig:lensing3}}
\hspace{1cm}
\subfigure[Retrograde Light trajectories in the rotating naked singularity for $M=1, a=0.8$]
{\includegraphics[width=80mm]{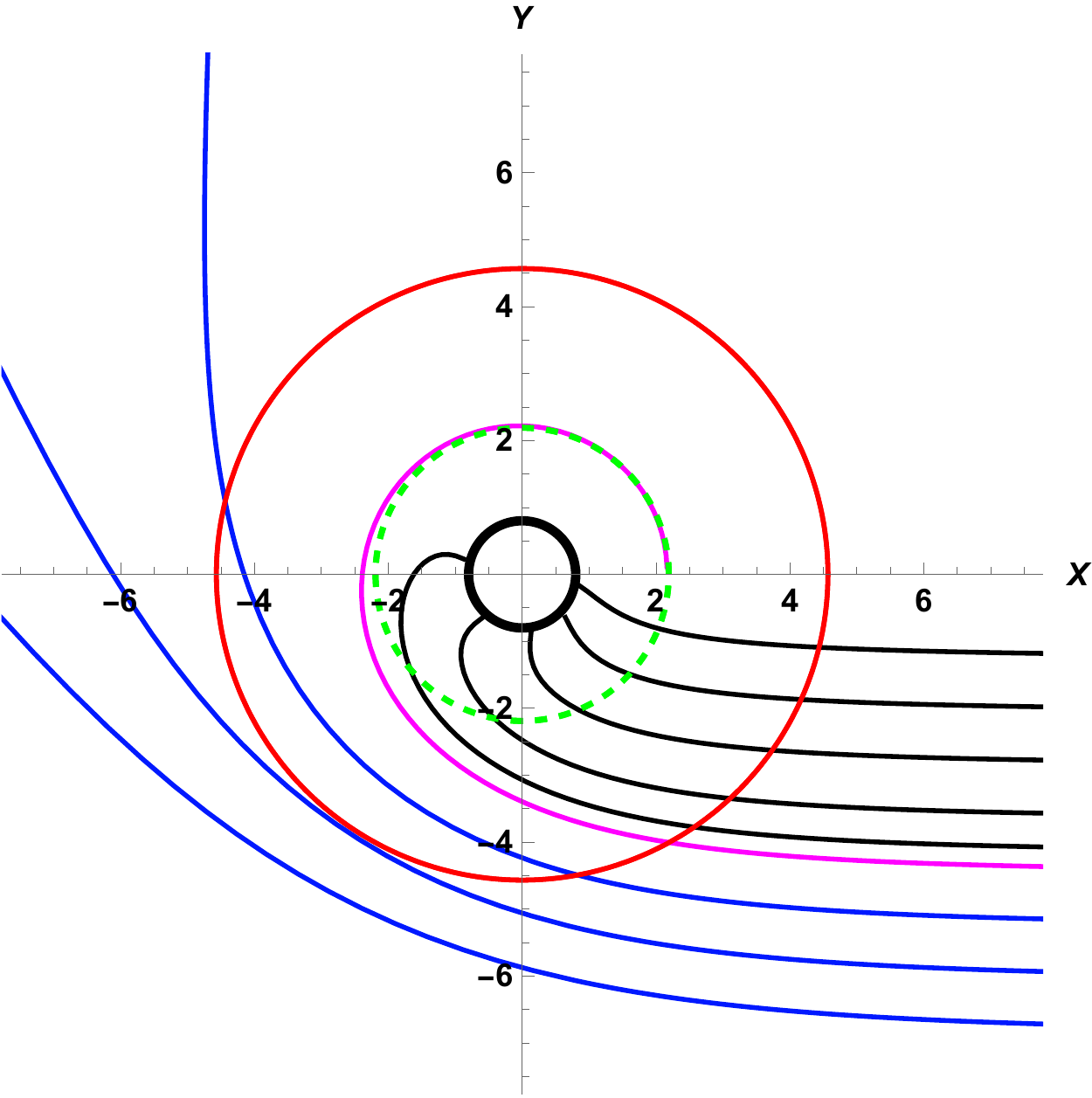}\label{fig:lensing4}}
\hspace{1cm}
 \caption{Figs.\,(\ref{fig:lensing1}) and (\ref{fig:lensing2}) show the lensing of null trajectories in the equatorial plane of a RNS and Kerr black hole, respectively. In this figure, the spin axes of the compact objects point out of the page. Here, the red solid circle and green dashed circle denote the critical impact parameters and photon spheres, respectively. The scattered light rays are represented by blue, and those plunging into the compact object are denoted by black. The light rays denoted by pink correspond to the critical impact parameter. In Figs.\,(\ref{fig:lensing1}- \ref{fig:lensing2}), the black shaded regions at the centre represent the outer event horizon of the Kerr black hole. Whereas in Figs.\,(\ref{fig:lensing3}- \ref{fig:lensing4}), the central black rings correspond to the ring singularity.}\label{fig:Lensing}
\end{figure*}
\begin{figure*}
   \centering
\subfigure[ shadow shape for the rotating naked singularity]
{\includegraphics[width=85mm]{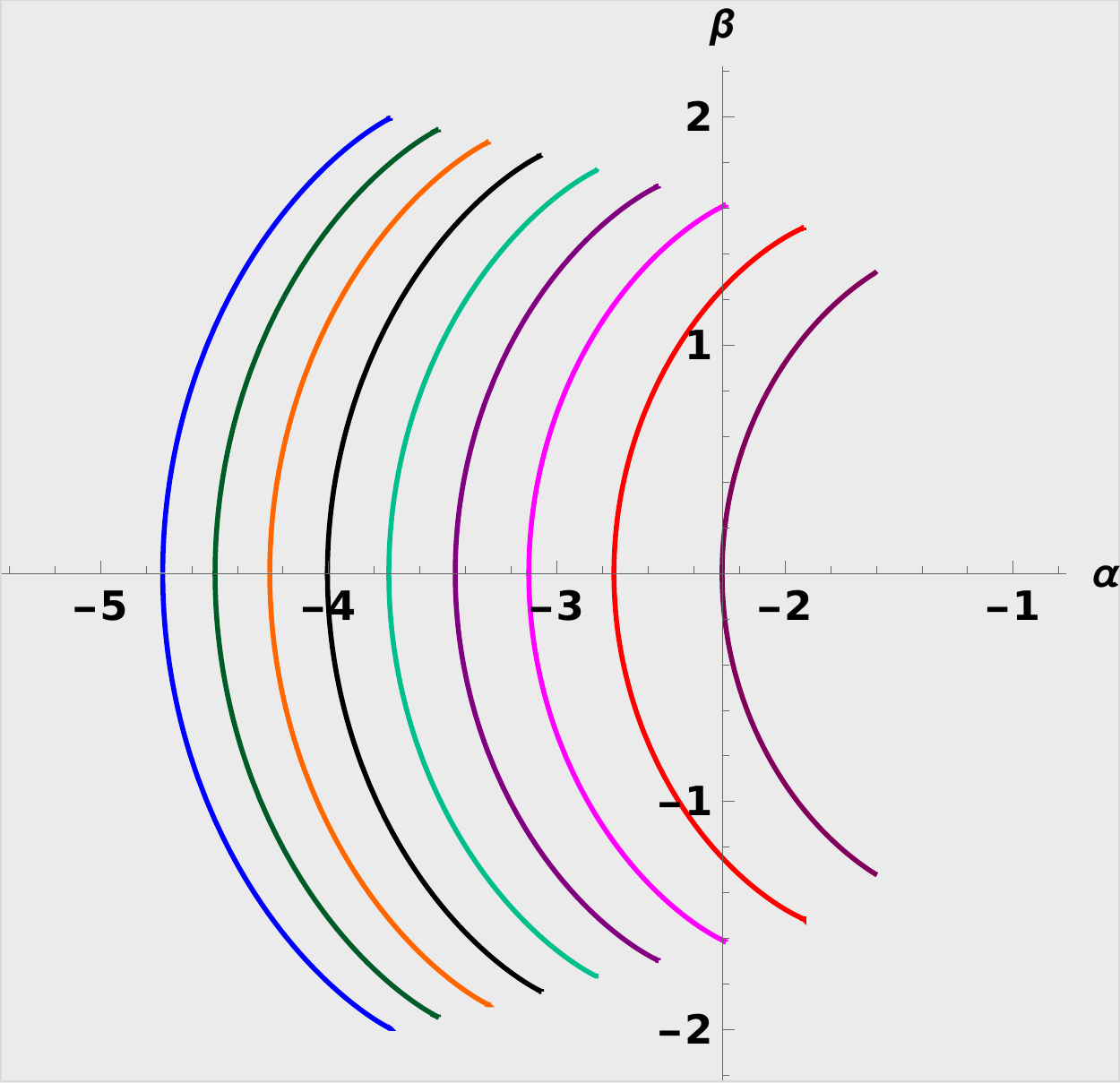}\label{fig:null singularity shadow shape}}
\hspace{0.5cm}
\subfigure[shadow shape for the Kerr Black hole]
{\includegraphics[width=83mm]{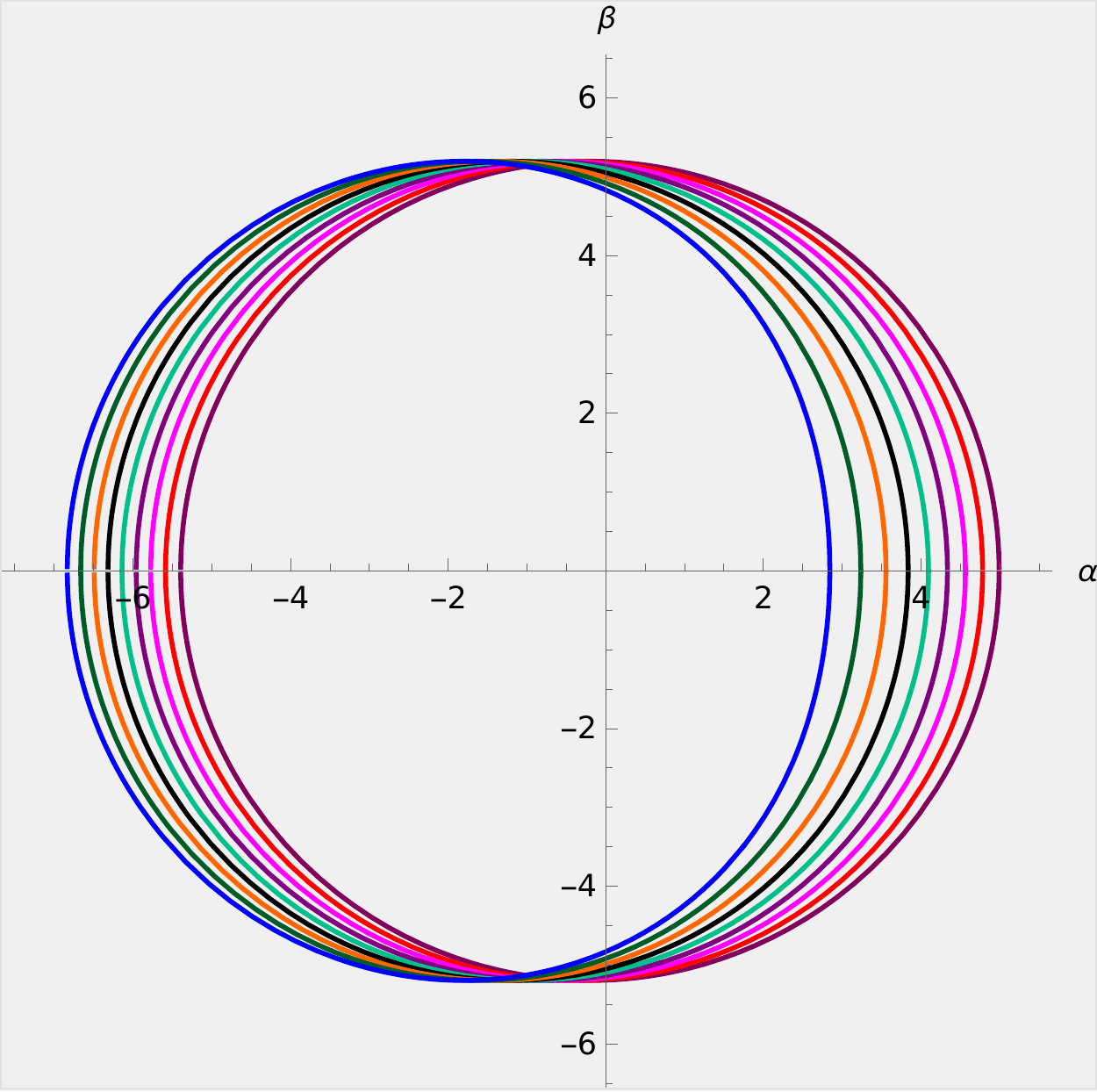}\label{fig:kerr shadow shape}}

 \caption{The figure shows the shadow outline cast by RNS and Kerr black hole. For both cases, the value of mass is set to one. The shadow shape indicated by blue to magenta color lines is formed for the spin parameters 0.9 to 0.1. The detailed discussion of this shadow images is given in section -\,(\ref{sec_discussion}).}
 \label{fig:shadow shape}
\end{figure*}

\begin{equation}
    \frac{1}{\Delta} ((r^2 + a^2)\,\epsilon - a\, L)^2 - (L - a\, \epsilon)^2 + \Delta \left(\frac{d S_r(r)}{d r}\right)^2  = K\,.
    \label{r part}
\end{equation}
In above differential equation, $K$ is the separation constant. Eqs.\,(\ref{theta part}) and (\ref{r part}) can be considered as a part of effective potential where the tangential part  and the radial part of the effective potential can be written as,    
\begin{eqnarray}
  R(r) &=& ((r^2 + a^2) - a \xi)^2 - \Delta (\eta + (\xi - a)^2)\,, \label{Rr}\\
 \Theta(\theta) &=& \eta - (\xi^2 cosec^2\theta - a^2 )\, cos^2\theta
    \label{thetatheta} \,.
\end{eqnarray}
 In Eq.\,(\ref{Rr}), $\xi = \frac{L}{\epsilon}$ and $ \eta = \frac{K}{\epsilon^2}$.
 In the equatorial plane, $\xi$ is the impact parameter of light geodesics and it can be used to distinguish the null geodesics.  Note that for a motion of photon, the $\Theta(\theta)$ and $R(r)$ should be positive. Using the expressions of $\Theta(\theta)$ and $R(r)$, one can characterize certain critical values of parameters $\xi~, \eta$ and under small perturbations, these critical values cause escape or plunge orbit for photons. On that count, we mark that the outline of the shadow is majorly dependent on the critical impact parameters. The standard conditions to determine the unstable circular orbit of photons are,  
\begin{equation}
     R(r_{ph}) = 0, \;   \frac{d R(r_{ph})}{dr}=0, \;    \frac{d^2 R(r_{ph})}{dr^2} \leq 0\,\,. \label{eq3}
\end{equation}
where, $r_{ph}$ is the radius of the photon sphere. Using the conditions (\ref{eq3}), one can obtain the critical impact parameters corresponding to the maxima of the $R(r)$ for RNS and Kerr Black hole spacetimes as,\,
\begin{eqnarray}
    && \eta_{RNS} = \frac{-r^6 + 4 a^2 M (M + r)^3}{a^2 (r + 2 M)^2},\label{eta_null}
    \\
    &&\xi_{RNS} = \frac{- 2 a^2 M^3 - r ( 6 a^2 M^2  + 4 a^2 M r + a^2 r^2 + r^4 )}{a r^2 (r + 2 M)}\,.\nonumber\\
\label{xi_null}
\end{eqnarray}
and 
\begin{eqnarray}
    && \eta_{Kerr} =\frac{r^3 \left(4 a^2 M-r (r-3 M)^2\right)}{a^2 (M-r)^2},\label{eta_kerr}
    \\
    &&\xi_{Kerr} =\frac{a^2 M+a^2 r-3 M r^2+r^3}{a (M-r)}\,.
\label{xi_kerr}
\end{eqnarray}

Since $\eta=0$ at the equatorial plane, the roots of Eqs.\,(\ref{eta_null}) and (\ref{eta_kerr}) determine the critical orbits of the photons~($r_{ph}$) in RNS and Kerr spacetimes, respectively at that plane. In Figs.\,(\ref{fig:eta_xi_null}) and (\ref{fig:eta_xi_kerr}), the behaviour of $\eta$ and $\xi$ as a function of $r$ is shown, where we consider $a=0.8$ and $M=1$. Orbits with a positive $\xi$ value represent the prograde equatorial motion of photons, and those with a negative $\xi$ value depict the retrograde equatorial orbit. For RNS, $\eta_{RNS}$ vanishes at $r=2.0389$, as shown in Fig.~(\ref{fig:eta_xi_null}). The circular orbit corresponding to this radius is retrograde, since $\xi_{RNS}$ is negative at $r_{ph}=2.0389$, and we observe that no photon sphere exists for prograde equatorial motion of photons. Hence, there is only one photon sphere at $r_{ph}=2.0389$ for the retrograde motion of null geodesics in RNS spacetime, and the corresponding critical impact parameter is $\xi_{RNS}(r_{ph})=4.49759$. On the other hand, in Fig.~(\ref{fig:eta_xi_kerr}), we see that there are three roots of $\eta_{Kerr}$. As we are only interested in orbits outside the event horizon, the first root $r_1$ must be discarded, since $r_1< r_{+}$. Here, $r_{ph+}$ is associated with the photon orbit having $\xi_{Kerr}>0$. Therefore, it represents the prograde equatorial orbit of a photon. In the same way, $r_{ph-}$ corresponds to the retrograde equatorial orbit, as it is associated with $\xi_{Kerr}<0$.
At the equatorial plane, there is a critical impact parameter ($\xi_{ph}$) corresponding to the unstable circular photon orbits $(r_{ph})$ below which the incoming null geodesics will plunge into the central compact object. The null geodesics with impact parameters larger than $\xi_{ph}$ will be scattered away, while those with impact parameters equal to $\xi_{ph}$ will remain trapped in the photon sphere at $r=r_{ph}$. Since the photon sphere corresponds to the unstable photon orbits, any small perturbation in the $r$-direction would cause the photons at $r=r_{ph}$ to fall into the singularity or scatter to infinity. Only the scattered photons from the compact object can reach the observer at infinity and create bright patches in the observer's sky. On the other hand, photons falling into a compact object do not reach the observer at infinity, resulting in a dark region. The shadow of the compact object is created by the combination of light and dark spots in the observer's sky. 

Figs.\,(\ref{fig:lensing1}, \ref{fig:lensing2}) and Figs.\,(\ref{fig:lensing3}, \ref{fig:lensing4}) depict the deflection of light trajectories in the equatorial plane of Kerr black hole and a RNS respectively with $a=0.8$ and $M=1$. In these figures, both the compact objects have spins in counter-clockwise direction. As shown in Fig.\,(\ref{fig:lensing3}), there is no photon sphere for prograde light trajectories; therefore, the light rays are dispersed away from the singularity. However, due to the finite size of the ring singularity in the equatorial plane, if the impact parameter of the light geodesics is smaller than the coordinate radius $a$ of the singularity, the light geodesics will become discontinuous as they approach the singularity and never reach the observer. For retrograde orbits of light geodesics, there exists a photon sphere in the equatorial plane, which is shown in Fig.~(\ref{fig:lensing4}) by a green dashed circle. Due to the presence of the photon sphere, light rays with impact parameters smaller than the critical impact parameter are trapped inside the photon sphere, creating dark patches in the observer's sky. Here, $\xi_{RNS}(r_{ph})=4.49759$ represents the radius of the shadow in the equatorial plane.

As explained earlier, the parameters $\eta$ and $\xi$ are essential to finding the shape of the shadow. If we consider the celestial coordinates $\alpha$ and $\beta$, which are the coordinates of the asymptotic observer sky, then using these celestial coordinates, one can obtain the apparent shape of the shadow seen by the asymptotic observer. The general expression to derive celestial coordinates $\alpha$ and $\beta$ are,
\begin{eqnarray}
    \alpha &=& \lim_{r_0\to\infty} \left(-r_0^2 \sin{\theta_{0}} \frac{d \phi}{dr}\Big\vert_{(r_o,\theta_o)}\right)
    \label{al} ,\\
 \beta &=& \lim_{r_0\to\infty} \left(r_0^2 \frac{d \theta}{dr}\Big\vert_{(r_o,\theta_o)}\right) \label{be},
\end{eqnarray}
where $r_0$ and $\theta_0$ are the coordinates of the asymptotic observer. Here, we consider asymptotically flat rotating metric. Therefore, in the RNS and Kerr black hole spacetimes, the expressions (\ref{al}) and (\ref{be}) will take the form as,
\begin{eqnarray}
    \alpha &=& \xi \, cosec\,\theta_0\,\, ,\\
    \beta &=& \pm \sqrt{\eta + a^2 \cos^2{\theta_{0}} - \xi^2\cot^2{\theta_{0}}}\,\, ,
\end{eqnarray}
By substituting Eqs.\,(\ref{eta_kerr}) and (\ref{xi_kerr}) for Kerr black hole, and Eqs.\,(\ref{eta_null}) and (\ref{xi_null}) for RNS spacetime in above expressions, one can deduce the corresponding celestial coordinates $\alpha$ and $\beta$ in Kerr and RNS spacetimes.
We construct the outline of the shadows using the radius of unstable circular orbit in the expressions of celestial coordinates for RNS and Kerr black hole spacetimes, where we have considered the inclination angle of the observer $\theta= \pi/2$. 
Fig.\,(\ref{fig:shadow shape}) represents the apparent size of the shadow cast by RNS and Kerr black hole for different values of spin parameters.  
The analysis of shadow shape is discussed in the next section. 
\section{Discussion and Conclusion}
 \label{sec_discussion}
In this paper, we carry out a comparative study of the properties of the lightlike geodesics and shadow in the rotating naked singularity~(RNS) and Kerr black hole spacetimes. First, we obtain the stationary and axisymmetric rotating null singularity spacetime by applying the Newman-Janis algorithm without the complexification process from the static and spherically symmetric null naked singularity~(NNS) spacetime. The orbit equation for null geodesics is then derived in order to plot the light trajectories around RNS and Kerr black holes. From the Hamilton-Jacobi formalism, we obtain the separable solution of radial~$(R(r))$ and azimuthal~$(\Theta(\theta))$ parts of the effective potential. Using the conditions of unstable circular orbits, we have studied the equatorial prograde and retrograde photon orbits.
The following are the outcomes of this paper: 
\begin{itemize}

    \item In the Kerr black hole, there are two horizons and two stationary limit surfaces. However, no such horizons or stationary limit surfaces are present in the RNS spacetime. Therefore, the ergoregion does not exist in an RNS. 

  \item We have studied the properties like nature of singularity, light trajectory, and shadow shape due to the photon sphere of RNS. As per the given spacetime in \cite{Joshi2020}, the static null naked singularity spacetime contains the null like singularity. Whereas, due to the inclusion of non-zero spin, the RNS spacetime has the time like singularity. Furthermore, there is no photon sphere for static null singularity \cite{Joshi2020}. In the case of RNS spacetime, however, a photon sphere exists for the retrograde motion of the null geodesics. 

     \item In Fig.\,(\ref{fig:eta_xi_kerr}), we can see that there are two radii of the photon sphere in the equatorial plane of the Kerr black hole. One radius corresponds to the prograde motion of light geodesics, which has a positive value of $\xi$ when $\eta$ = 0. Another radius is for a retrograde motion of light, where the value of $\xi$ is negative when $\eta$ = 0. Besides that, for an RNS in Fig.\,(\ref{fig:eta_xi_null}), there exists only one photon sphere in the equatorial plane, and that is for retrograde motion of light geodesics. As a result, in the case of an RNS, all equatorial light rays travelling in prograde motion will be scattered by the central singularity. While for the retrograde light orbits, light geodesics with impact parameters smaller than the critical impact parameter will fall into the singularity. This can be understood from the behavior of light trajectories shown in Fig.\,(\ref{fig:Lensing}).
     
\item We know that the rotating gravitational sources drag the surrounding spacetime. This dragging effect can be observed from the deflection of light geodesics in Fig.\,(\ref{fig:Lensing}). From Fig.\,(\ref{fig:lensing2}), it is apparent that the dragging effect is largest near the outer event horizon of the Kerr black hole, causing the retrograde light orbits to change their direction of motion in the direction of the spin of the Kerr black hole. Whereas for RNS spacetime, the dragging effect is most significant around the ring singularity, as seen in Fig.\,(\ref{fig:Lensing}).
 
 \item The RNS casts an arc shape shadow which can be seen in Fig.\,(\ref{fig:null singularity shadow shape}). That is because of the particular occurrence of the photon sphere around RNS. The shadow in Fig.\,(\ref{fig:kerr shadow shape}) is cast by Kerr black hole. In both cases, shadow shape is shown for different spin parameters indicated by blue to magenta color lines are formed for the spin parameters 0.9 to 0.1, respectively. The shadow cast by RNS is almost half of that of the Kerr black hole. With the increasing spin, the shadow size in RNS increases, whereas in Kerr black hole, the deformation in the shadow shape increases.
  
In this work, we restricted ourselves to examining the shadow shape due to the presence of the photon sphere in RNS spacetime. In general, one may show that in the absence of photon sphere, the shadow can be cast by a rotating compact object due to the existence of the ring singularity.
\end{itemize}

\acknowledgments
V. P. would like to acknowledge the support of the SHODH fellowship (ScHeme Of Developing High quality research - MYSY).

\end{document}